\documentclass[twocolumn,amsmath,amssymb]{revtex4}
\usepackage{bm}
\usepackage{graphicx}
\usepackage{epstopdf}

\begin{document}

\newcommand{\pp}[1]{\phantom{#1}}
\newcommand{\be}{\begin{eqnarray}}
\newcommand{\ee}{\end{eqnarray}}
\newcommand{\Sinn}{\textrm{Sin }}
\newcommand{\Coss}{\textrm{Cos }}
\newcommand{\Sin}{\textrm{Sin}}
\newcommand{\Cos}{\textrm{Cos}}
\newcommand{\arcsinh}{\textrm{arcsinh }}

\title{
Contra Bellum: Bell's theorem as a confusion of languages 
}
\author{Marek Czachor}
\affiliation{
Instytut Fizyki i Informatyki Stosowanej,
Politechnika Gda\'nska, 80-233 Gda\'nsk, Poland
}
\begin{abstract}
Bell's theorem   is a conflict of mathematical predictions formulated within an infinite hierarchy of mathematical models. Inequalities formulated at level $k\in\mathbb{Z}$, are violated by probabilities at level $k+1$. We are inclined to think that $k=0$ corresponds to the the classical world, while the quantum one is $k=1$. However, as the $k=0$ inequalities are violated by $k=1$ probabilities, the same relation holds between $k=1$ inequalities violated by $k=2$ probabilities, $k=-1$ inequalities, violated by $k=0$ probabilities, and so forth. Accepting the logic of the Bell theorem, can we prove by induction that nothing exists? 
\end{abstract}

\maketitle

\section{Is Bell's theorem a mathematical theorem?}

If we treat Bell's theorem \cite{Bell} as a theorem about  additivity of Lebesgue measures,  then yes: this is a mathematical theorem. However, Bell's theorem is more ambitious. It tells us about reality per se,  security of communication channels,  structure of space and time, or even freedom of experimental physicists...

Although mathematical theorems cannot have counterexamples, this is not necessarily true for theorems about physical reality. The whole history of science is a series of exceptions to various well established truths.

One such famous truth about reality was known as Euclid's fifth axiom, essentially stating that angles in any triangle add up to $180^\circ$. It was so  self-evident to 19th century mathematicians that even Gauss himself was not eager to publish his thoughts on the subject. 

Bell's theorem is technically based on another apparently  self-evident truth about additivity, namely
\be
\int_\Lambda (f\pm g)(\lambda)d\lambda
=
\int_\Lambda f(\lambda)d\lambda
\pm
\int_\Lambda g(\lambda)d\lambda.\label{1}
\ee
In proofs of Bell-type inequalities one often replaces (\ref{1}) by a more elementary rule,
\be
\frac{n\pm m}{N}=\frac{n}{N}\pm \frac{m}{N}.\label{2}
\ee
(\ref{1}) occurs in contexts of  probability measures, while (\ref{2}) is typical of frequentists approaches. 

However, neither (\ref{1}) nor (\ref{2}) are universally true. (\ref{1}) fails for fuzzy or fractal functions. 
(\ref{2}) fails if $n$, $m$, represent velocities and $N$ is the velocity of light. In the latter case, what we rather get is
\be
n\oplus m &=&
N\tanh\left(\tanh^{-1}\frac{n}{N}+\tanh^{-1}\frac{m}{N}\right)\label{3}\\
&=&
f^{-1}\big(f(n)+f(m)\big).
\label{3'}
\ee
Of course, nothing can prevent us from adding velocities by means of (\ref{2}), but this is not what Nature does. The arithmetic of Nature is (\ref{3}).
A relation between $\oplus$ and $+$ is here analogous to the one between curvature of  a general manifold and flatness of its local chart of coordinates (charts in atlases are flat).
Arithmetic in special relativity becomes as physical as geometry in general relativity.

A similar situation occurs with (\ref{1}). In fuzzy and fractal applications one often encounters
\cite{fuzzy dif,Schmeidler,Mesiar1995,fuzzy calc,Marinacci,Pap2002,G,ACK2016b,BC}
\be
\int_\Lambda f(\lambda)d\lambda
\oplus 
\int_\Lambda g(\lambda)d\lambda
=
\int_\Lambda (f\oplus g)(\lambda)d\lambda
.\label{4}
\ee
The exact meaning of $\oplus$ depends on the way fuzzy sets are constructed \cite{LeiXu}, or which fractals one is dealing with \cite{ACK2018}. A particular example of (\ref{4}) occurs  in Maslov's idempotent analysis \cite{Maslov1,Maslov2}. Here, certain optimization problems that are non-linear in the usual framework become linear with respect to generalized arithmetic operations \cite{Gondran}, even though the generalized arithmetic is not isomorphic to the one of $\mathbb{R}$ \cite{LMS}.

The goal of the paper is to show that quantum probabilities typical of a two-particle singlet state (that is, those used by Bell in his argumentation), despite all the wisdom of theoretical physicists, {\it can\/} result from a local theory where Einstein-Podolsky-Rosen-type elements of reality exist \cite{EPR,Bohm}, with probabilities given by local realistic Clauser-Horne formulas \cite{CH}, where observers have free will, and their detectors are 100\% efficient. The only difference is in the form of the integral, whose  linearity is  with respect to $\oplus$, $\ominus$, $\odot$ and $\oslash$ appropriately defined.

Violation of Bell-type inequalities is then no more paradoxical than $c+c=2c$ which could be claimed to violate the speed of light limit. Moreover, there is no problem with circumventing the Tsirelson bounds typical of Hilbert-space models of probability \cite{Tsirelson}, still maintaining Bell locality, EPR elements of reality, free will of observers, and 100\% efficient detectors. 

I can reassure the readers that the models we are analyzing are {\it not\/} an alternative to quantum mechanics. They do not explain why probability amplitudes interfere, but nevertheless shift the discussion of linearity to new, unexplored areas. The status of theorems based on algebraic properties of observables, such as the  Greenberger-Horne-Zeilinger theorem \cite{GHSZ,GHZ} or its single-particle analogues  \cite{MC1994}, is still open. 

But what the models {\it do\/} show is that quantum correlations of an EPR type do not necessary exclude EPR elements of reality, a conclusion with potentially dramatic implications for quantum cryptography. How serious the consequences are remains to be investigated. Einstein's views on incompleteness of quantum mechanics receive unexpected support.

We will begin our discussion with the observation that principles of relativity are more general and ubiquitous than Einstein's relativity of uniform motion, or Copernican relativity of point of observation. The most fundamental principle, occurring in all natural sciences, is relativity of arithmetic \cite{MC2015}. It implies, in particular,  principles of relativity of calculus and relativity of probability. 

Both are essential for Bell's theorem. 

\section{Relativity of probability}

Relativity of probability occurs at several levels. The most obvious one is illustrated by the following example. It can be regarded as a particular case of  Einstein's special relativity.

Assume a source emits particles to the right or to the left, with certain probability density $\rho(v)$ of velocities. If $N$ particles have been emitted, let $N_+$ denote the number of particles propagating to the right. An observer measures $N_+/N$ and compares it with the theoretical prediction,
$p_+=\int_0^\infty \rho(v)dv$. 

An observer that moves with velocity $V$ with respect to the previous one will measure a different value of $N_+/N$, even though  both of them analyze the same experiment, with the same $N$.

The example is trivial, but it illustrates an important fact about probability: Different observers may associate different probabilities with the same experimental situation, and with the same definition of elementary events. In this concrete example the relativity of probability, $p_+(V)=\int_0^\infty \rho(v+V)dv$, results from relativity of motion.

As a less trivial relativistic example consider gravitational collapse of a star. There are two observers: Alice who falls with the star, and Bob who remains at rest at position $r$. Alice employs a broken clock that randomly fails to work (which happens with probability $p_0$). The motion of the clock's hand becomes a Poisson process characterized by probability $p_1=1-p_0$ of a forward move. 

A single run of experiment lasts a fixed amount $\tau$ of observer's proper time.
Alice measures $N=\lfloor\tau/\Delta \tau\rfloor$ bits $A_1,\dots,A_N$ ($N_1$ events $A_j=1$ when the clock's hand moves; $N_0=N-N_1$ events $A_j=0$ when it gets stuck). The average amount of proper time measured by the damaged clock of Alice is $p_1\tau$.

The experimental ratio $N_1/N$ observed by Alice gets translated into 
$\tilde N_1/\tilde N$
observed by Bob. In general, $\tilde N_1\neq N_1$ and $\tilde N\neq N$ because numbers of observed events differ for Alice and Bob due to relativity of time and the presence of the horizon. The events observed by Alice after she will have crossed the Schwarzschild radius at her proper  time $\tau_S$ will be unavailable to Bob, even though his detectors are 100\% efficient. 

Bob should cautiously draw conclusions about $N_1$ and $N$ on the basis of $\tilde N_1$ and $\tilde N$ he observes. 
For example, if he concludes that $\tau_S$ is greater than $\tau$ because from his perspective Alice cannot reach the Schwarzschild radius, this inequality can be `violated' in the world of Alice.

Bob can derive various inequalities about the data of Alice, provided he knows the map $g_r$ that relates her $N_1/N$ with his
\be
\tilde N_1/\tilde N
=g_r(N_1/N).\label{5}
\ee
The exact form of $ g_r $ is irrelevant for our argument, but it could be derived on the basis of general relativity, if needed.
 
From our perspective, it is important that $ g_r $ connects two real probabilistic processes. Both $ N_1/N $ and $ g_r (N_1/ N) $ are true, physically significant probabilities. The `violation' of 
Bob's world $\tau_S>\tau$ by Alice's world $\tau_S<\tau$ is paradoxical only for those who do not understand Einstein's theory of gravity. 

\section{A lemma on relativity of binary probabilities}

For binary events there exists a simple result guaranteeing that both $N_1/N$ and $g(N_1/N)$ are probabilities.

{\it Lemma~1\/}: 
$g(p)+g(1-p)=1$, for any $p\in [0,1]$,  if and only if  
\be
g(p)=\frac{1}{2} + h\left(p-\frac{1}{2}\right),\label{6}
\ee
where $h(-x)=-h(x)$. Any such $g$ has a fixed point at $p=1/2$.

So, any antisymmetric $h(x)$ leads to an acceptable $g(p)$. The proof can be found in \cite{Czachor2021}. For a discussion of non-binary probabilities see \cite{Entropy}.

As an example consider the antisymmetric function
\be
h(x)
=
\frac{1}{2}\sin \pi x.\label{7}
\ee
Then  $g(p)=\sin^2\frac{\pi}{2}p$, and indeed
\be
g(p)+g(1-p)=\sin^2\frac{\pi}{2}p+\cos^2\frac{\pi}{2}p=1
\ee
for any $p$. 
Now let $p=(\pi-\theta)/\pi$ be the probability of finding a point belonging to the overlap of two half-circles rotated by 
$\theta$. Then
\be
g(p)=\sin^2\frac{\pi}{2}\frac{\pi-\theta}{\pi}=\cos^2\frac{\theta}{2}\label{cos}
\ee
is the Malus law for spin 1/2 (or Mach-Zehnder interferometers).

Note that $g$ is one-to-one on $[0,1]$. Moreover, $g(0)=0$ and $g(1)=1$, a property with important implications for definition of bits: classical, quantum, and intermediate. 

The readers should think of $p$ and $\tilde p=g(p)$ in categories similar to those that have led us to  (\ref{5}). Both $p$ and $\tilde p$ can be physically meaningful. We should be as cautious as Bob in formulating statements about the level of $p$ on the basis of the rules that apply to the level of $\tilde p$.

\section{Arithmetic elements of reality}

Consider some set $\mathbb{X}$ and a bijection $f_\mathbb{X}:\mathbb{X} \to\mathbb{R}$. Cardinality of $\mathbb{X}$  must be the same as the one of $\mathbb{R}$.
The inverse map  is $g_\mathbb{X}=f_\mathbb{X}^{-1}$, $g_\mathbb{X}:\mathbb{R} \to\mathbb{X}$. The map $g$ from the previous section can be an example of  $g_\mathbb{R}$ restricted to $[0,1]$. Putting it differently, the bijection $g:[0,1]\to [0,1]$ can be extended to a bijection $g_\mathbb{R}:\mathbb{R} \to\mathbb{R}$, satisfying $g_\mathbb{R}(p)=\sin^2\frac{\pi}{2}p=g(p)$ when restricted to $p\in[0,1]$. 

We define arithmetic operations in $\mathbb{X}$,
\be
x\oplus_\mathbb{X} y  &=& g_\mathbb{X}\big( f_\mathbb{X}(x)+f_\mathbb{X}(y)\big),\label{X1}\\
x\ominus_\mathbb{X} y  &=& g_\mathbb{X}\big( f_\mathbb{X}(x)-f_\mathbb{X}(y)\big),\label{X2}\\
x\odot_\mathbb{X} y  &=& g_\mathbb{X}\big( f_\mathbb{X}(x)\cdot f_\mathbb{X}(y)\big),\label{X3}\\
x\oslash_\mathbb{X} y  &=& g_\mathbb{X}\big( f_\mathbb{X}(x)/f_\mathbb{X}(y)\big).\label{X4}
\ee
The arithmetic given by (\ref{X1})--(\ref{X4}) is called projective \cite{BC,Burgin2010}. $f_\mathbb{X}$ defines an isomorphism of arithmetics. The neutral elements, $0_\mathbb{X}=g_\mathbb{X}(0)$ (projective zero in $\mathbb{X}$), $1_\mathbb{X}=g_\mathbb{X}(1)$ (projective one in $\mathbb{X}$) are to some extent analogous to qubits \cite{Czachor2020,Entropy}.

Indeed, expressions such as $0_\mathbb{X}+0_\mathbb{Y}$ are in general meaningless if $\mathbb{X}\neq \mathbb{Y}$. Just think of $\mathbb{X}=\mathbb{R}$ and $\mathbb{Y}=\mathbb{R}^2$. 
Even if $\mathbb{X}=\mathbb{Y}$ and $0_\mathbb{X}=0_\mathbb{Y}=0$, $1_\mathbb{X}=1_\mathbb{Y}=1$, the projective bits can be as incompatible as eigenvalues of non-commuting projectors.

However, in spite of this incompatibility, $0_\mathbb{X}=g_\mathbb{X}(0)$ and $0_\mathbb{Y}=g_\mathbb{Y}(0)$ are  images of the same $0\in \mathbb{R}$. This `ordinary zero' can play the role of an EPR-type element of reality for $0_\mathbb{X}$ and $0_\mathbb{Y}$:
Incompatible projective bits can be correlated by means of their elements of reality, in exact analogy to the formulas postulated by Bell in his classic analysis. 

Note that (\ref{3}) is an example of (\ref{X1}). The neutral elements are: $0_\mathbb{X}=N\tanh 0=0$, $1_\mathbb{X}=N\tanh 1=0.76 N$ (hence velocity $0.76\,c$ is the neutral element of special relativistic multiplication). Velocity of light is literally infinite, of course in the sense of $\infty_\mathbb{X}=N \tanh(\infty)=N$.
$c\oplus c=c$ is an example of $\infty_\mathbb{X}\oplus_\mathbb{X}\infty_\mathbb{X}=\infty_\mathbb{X}$. Strictly speaking, a relativistic unit of velocity is not $c$ but $c \tanh 1$.

\section{Clauser-Horne formulas for projective bits}

We are interested in singlet-state probabilities,
\be
P_{0_1} &=& P_{1_1} =P_{0_2} = P_{1_2}\nonumber\\
&=&
\langle\psi|\hat P_{0_1}\otimes I|\psi\rangle=\langle\psi|\hat P_{1_1}\otimes I|\psi\rangle\nonumber\\
&=&
\langle\psi|I\otimes \hat P_{0_2}|\psi\rangle=\langle\psi|I\otimes \hat P_{1_2}|\psi\rangle=\frac{1}{2},\label{14}
\ee
with joint probabilities,
\be
P_{0_10_2} &=& P_{1_11_2} 
=\langle\psi|\hat P_{0_1}\otimes \hat P_{0_2}|\psi\rangle
=\langle\psi|\hat P_{1_1}\otimes \hat P_{1_2}|\psi\rangle
\nonumber\\
&=&\frac{1}{2}\sin^2\frac{\alpha-\beta}{2},\\
P_{0_11_2} &=& P_{1_10_2} 
=\langle\psi|\hat P_{0_1}\otimes \hat P_{1_2}|\psi\rangle
=\langle\psi|\hat P_{1_1}\otimes \hat P_{0_2}|\psi\rangle\nonumber\\
&=&\frac{1}{2}\cos^2\frac{\alpha-\beta}{2}.
\ee
We will write them in a Clauser-Horne form \cite{CH}
\be
P_{A_1A_2} &=& \int \chi_{A_1}(x)\odot_\mathbb{X} \chi_{A_2}(x)\odot_\mathbb{X} \rho(x){\rm D}x\label{16},\\
P_{A} &=& \int \chi_{A}(x)\odot_\mathbb{X} \rho(x){\rm D}x=\frac{1}{2},\label{17}
\ee
where the $\chi$s are characteristic functions and $\rho(x)\geq 0$ is a non-negative probability density normalized to 1,
\be
\int \rho(x){\rm D}x=1.
\ee
Of course, the trick is to work with appropriate forms of the integral, and employ the freedom available in possible meanings of $\odot_\mathbb{X}$ and $\oplus_\mathbb{X}$. We will assume $\mathbb{X}=\mathbb{R}$, and $g_\mathbb{X}(0)=0$, $g_\mathbb{X}(1)=1$. 
The latter two conditions imply that the values of projective bits will be given by ordinary 0 and 1.

Formulas  (\ref{16}) and (\ref{17}) implicitly imply that measurements are modeled in the usual way by products of  $\rho(x)$ with characteristic functions,
\be
\rho(x) &\mapsto& \chi_{A}(x)\odot_\mathbb{X} \rho(x),\\
\rho(x) &\mapsto& \chi_{A\cap B}(x)\odot_\mathbb{X} \rho(x)\\
&=&\chi_{A}(x)\odot_\mathbb{X} \chi_{B}(x)\odot_\mathbb{X} \rho(x),
\ee
and so forth. If $A'$ denotes the set-theoretic completion of the set $A$, then:
\be
\chi_{A}(x)\oplus_\mathbb{X} \chi_{A'}(x) &=& 1,\\
1\ominus_\mathbb{X} \chi_{A}(x) &=& \chi_{A'}(x),\\
\chi_{A}(x)\odot_\mathbb{X} \chi_{A'}(x) &=& 0,\\
\chi_{A}(x)\odot_\mathbb{X} \chi_{A}(x) &=&\chi_{A}(x),\\
\chi_{A'}(x)\odot_\mathbb{X} \chi_{A'}(x) &=&\chi_{A'}(x).
\ee
The probabilities must add up to 1 in the ordinary way,
\be
P_{0_10_2}+P_{0_11_2}+P_{1_10_2}+P_{1_11_2} =1,\label{18}
\ee
because this is how experimentalists will use them. 
 
On the other hand, the integral can be additive in a more general sense of (\ref{4}), similarly to fuzzy, fractal, or idempotent integrals. A dual form of normalization will be a consequence of such a generalized linearity,
\be
P_{0_10_2}\oplus_\mathbb{X} P_{0_11_2}\oplus_\mathbb{X} P_{1_10_2}\oplus_\mathbb{X} P_{1_11_2} =1.\label{19}
\ee
(\ref{18}) and (\ref{19}) must hold simultaneously for any $P_{A_1A_2}$, a condition which is not entirely trivial, but whose solution exists.

The choice of arithmetic will naturally define the  integral occurring in (\ref{16})--(\ref{17}). Historically the first construction of a  calculus based on projective arithmetic was given by Grossman and Katz in their 1972 book {\it Non-Newtonian Calculus\/} \cite{GK,NNC}. 

Bell published his paper in 1964. 

\section{Non-Newtonian calculus}

We need an integral because Clauser-Horne formulas involve integration. 
In fuzzy or fractal applications the usual strategy would be to define some measure on a fractal or fuzzy set, and only then  start to worry if the resulting integral is consistent with derivatives, typically defined by means of a completely different procedure than the one that has led to the integral. In effect, the fundamental theorem of calculus often becomes problematic \cite{LeiXu}. 

The approach that starts with arithmetic is much more systematic. One first defines a derivative by means of a formula which is a straightforward generalization of
\be
\frac{{\rm d}F(x)}{{\rm d}x}=\lim_{\delta\to 0}\frac{F(x+\delta)-F(x)}{\delta}.\label{28}
\ee
Then one demands that the integral be related with the derivative by means of the fundamental theorem of calculus. The notion of a measure appears automatically at the very end, once we know how to integrate. Knowing the measure, we know how to define probability.

The non-Newtonian derivative of a function $F:\mathbb{X}\to \mathbb{Y}$ depends on arithmetics of $\mathbb{X}$ and 
$\mathbb{Y}$. Denoting $\delta_\mathbb{X}=g_\mathbb{X}(\delta)$, $\delta_\mathbb{Y}=g_\mathbb{Y}(\delta)$, one defines 
\be
\frac{{\rm D} F(x)}{{\rm D}x}
&=&
\lim_{\delta\to 0}
\big(F(x\oplus_\mathbb{X}\delta_\mathbb{X})\ominus_\mathbb{Y}F(x)\big)\oslash_\mathbb{Y}\delta_\mathbb{Y},
\label{DF}
\ee
whose more practical form reads
\be
\frac{{\rm D} F(x)}{{\rm D}x}
&=&
g_\mathbb{Y}
\left(
\frac{{\rm d}\tilde F\big(f_\mathbb{X}(x)\big)}{{\rm d}f_\mathbb{X}(x)}
\right).\label{DF'}
\ee
The argument of $g_\mathbb{Y}$ in (\ref{DF'})  is the (Newtonian) derivative (\ref{28}) of $\tilde F$ defined by  the commutative diagram
\be
\begin{array}{rcl}
\mathbb{X}                & \stackrel{F}{\longrightarrow}       & \mathbb{Y}               \\
f_\mathbb{X}{\Big\downarrow}    &                                     & {\Big\uparrow}g_\mathbb{Y}  \\
\mathbb{R}                & \stackrel{\tilde F}{\longrightarrow}   & \mathbb{R}
\end{array}.\label{33}
\ee
Although (\ref{DF'}) makes non-Newtonian differentiation as simple as the Newtonian one, (\ref{DF}) reveals the logical structure behind  the derivative. For example, it explains why we find the generalized form of additivity,
\be
\frac{{\rm D} \big(F(x)\oplus_\mathbb{Y}G(x)\big)}{{\rm D}x}
&=&
\frac{{\rm D} F(x)}{{\rm D}x}
\oplus_\mathbb{Y}
\frac{{\rm D} G(x)}{{\rm D}x}
\ee
and the generalized Leibniz rule
\be
&{}&\frac{{\rm D} \big(F(x)\odot_\mathbb{Y}G(x)\big)}{{\rm D}x}
\nonumber\\
&{}&
\pp=
=
\frac{{\rm D} F(x)}{{\rm D}x}
\odot_\mathbb{Y}
G(x)
\oplus_\mathbb{Y}
F(x)
\odot_\mathbb{Y}
\frac{{\rm D} G(x)}{{\rm D}x}.
\ee
In order to define a non-Newtonian integral $\int_a^b F(x) {\rm D} x$ we demand its consistency with the derivative
(two fundamental theorems of calculus):
\be
\int_a^b
\frac{{\rm D} F(x)}{{\rm D} x} {\rm D} x
&=&
F(b)\ominus_\mathbb{Y} F(a),\\
\frac{{\rm D}}{{\rm D} x}
\int_a^x
F(y) {\rm D} y
&=&
F(x).
\ee
The result is
\be
\int_a^b
F(x) {\rm D} x
&=&
g_\mathbb{Y}
\left(
\int_{f_\mathbb{X}(a)}^{f_\mathbb{X}(b)}
\tilde F(r) {\rm d}r
\right).\label{integr}
\ee
The argument of $g_\mathbb{Y}$ in (\ref{integr})  is the (Newtonian, hence Lebesgue, Riemann, etc.) integral of $\tilde F$ defined by  the commutative diagram (\ref{33}).
Such an integral inherits additivity, 
\be
\int_a^b
F(x)\oplus_\mathbb{Y}G(x) {\rm D} x
=
\int_a^b
F(x) {\rm D} x
\oplus_\mathbb{Y}
\int_a^b
G(x) {\rm D} x,
\ee
and one-homogeneity (for a constant $F$),
\be
\int_a^b
F\odot_\mathbb{Y}G(x) {\rm D} x
=
F\odot_\mathbb{Y}
\int_a^b
G(x) {\rm D} x,
\ee
from the arithmetic that defines the derivative.

It should be  now rather clear why non-Newtonian hidden-variable models lead to Bell-type inequalities of basically the usual  form, but with the ordinary plus, minus, times, and divided replaced by $\oplus$, $\ominus$, $\odot$, and $\oslash$.

If one takes this subtlety into account, then quantum mechanical singlet-state  probabilities will {\it not\/} violate the Bell inequality --- not the one that can be derived for the hidden-variable model.

\section{Singlet-state probabilities}

It remains to construct the projective arithmetic (\ref{X1})--(\ref{X4}) that implies (\ref{14})--(\ref{17}) by means of the corresponding non-Newtonian integral (\ref{integr}). $g_\mathbb{X}$ will be constructed via an intermediate $g$, whose properties are described by the following consequence of Lemma~1:

{\it Lemma 2\/}: Consider four joint probabilities $p_{0_10_2}$, $p_{1_11_2}$, $p_{0_11_2}$, $p_{1_10_2}$, satisfying 
\be
\sum_{AB}p_{AB} &=& 1,\label{L2a}\\
\sum_{A}p_{AA_2} &=& \sum_{A}p_{A_1 A}=\frac{1}{2}.\label{L2b}
\ee
A sufficient condition for
\be
\sum_{AB}G(p_{AB}) &=& 1,\label{L2G}
\ee
for any $p_{AB}$ satisfying (\ref{L2a}), (\ref{L2b}),  
is given by $G(p)=\frac{1}{2}g(2p)$, where $g$ satisfies Lemma~1. Any such $G$ has a fixed point at $p=1/4$. 

The proof can be found in \cite{MCKN}.

\begin{figure}
\includegraphics[width=8 cm]{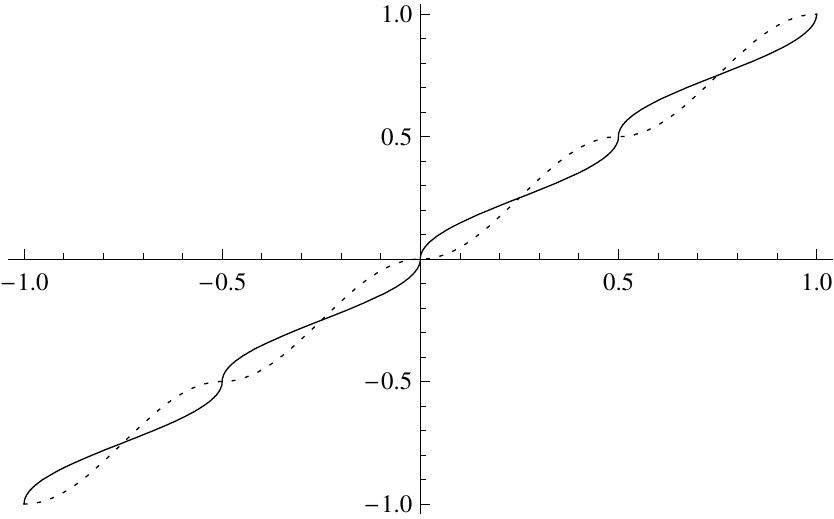}
\includegraphics[width=8 cm]{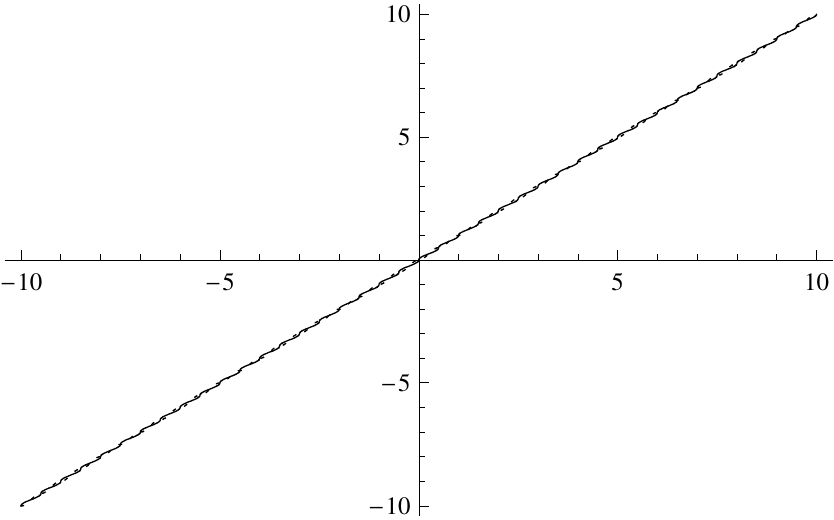}
\caption{One-to-one $f_\mathbb{X}:\mathbb{R}\to \mathbb{R}$  (full) and its inverse $g_\mathbb{X}$ (dotted) defined by  (\ref{f})  and (\ref{f^-1}), as implied by Lemma~2. Both functions have fixed points at integer multiples of $1/4$. The plots are given in two scales, $-1\le x\le 1$ and $-10\le x\le 10$, explaining the origin of the correspondence principle discussed in Sec.~\ref{Correspondence principles}.}
\label{Fig2}
\end{figure}

Guided by Lemmas 1 and 2, we take $\mathbb{X}=\mathbb{R}$ and define 
(Fig.~\ref{Fig2}),
\be
g_\mathbb{X}(x) &=&\frac{n}{2}+ \frac{1}{2}\sin^2\pi \left(x-\frac{n}{2}\right), \label{f^-1}\\
f_\mathbb{X}(x) &=& \frac{n}{2}+\frac{1}{\pi}\arcsin\sqrt{2x -n}, \label{f}\\
&\pp=&
\textrm{for $\frac{n}{2}\le x\le \frac{n+1}{2}$, $n\in\mathbb{Z}$},\nonumber
\ee
(for more details see \cite{Czachor2021}).
Function 
(\ref{f^-1}) is, up to the rescaling $g(p)\mapsto \frac{1}{2}g(2p)$ required by Lemma~2, the one we have used as the illustration of Lemma~1 for spins 1/2, but extended from $[0,1]$ to the whole of $\mathbb{R}$.
Non-Newtonian integrals (\ref{16})--(\ref{17}) constructed by means of (\ref{f^-1})--(\ref{f}) reconstruct singlet-state probabilities if we appropriately define the characteristic functions. 

For example,
\be
\left(\int_{\alpha_\mathbb{X}}^{\beta_\mathbb{X}} {\rm D}\lambda\right) \oslash_\mathbb{X}g_\mathbb{X}(2\pi)
=
\frac{1}{2}\sin^2\frac{\beta-\alpha}{2},\label{46'}
\ee
$\alpha_\mathbb{X}=g_\mathbb{X}(\alpha)$, $\beta_\mathbb{X}=g_\mathbb{X}(\beta)$, 
which is the standard local hidden-variable expression postulated by Bell. It can be rewritten as a particular case of  (\ref{16}) if we denote
\be
\rho(\lambda)
=
1\oslash_\mathbb{X}g_\mathbb{X}(2\pi)
=
g_\mathbb{X}\big(1/(2\pi)\big),
\ee
and integration is over the circle $0\le \lambda<g_\mathbb{X}(2\pi)$. The product of characteristic functions is encoded in the  integration limits.

Rotational invariance of the probability is a consequence of 
\be
\int_{\alpha_\mathbb{X}}^{\beta_\mathbb{X}} {\rm D}x
=
\int_{\alpha_\mathbb{X}\oplus_\mathbb{X}\gamma_\mathbb{X}}^{\beta_\mathbb{X}\oplus_\mathbb{X}\gamma_\mathbb{X}} {\rm D}x
\ee
valid for any $\gamma_\mathbb{X}\in \mathbb{X}$, and any non-Newtonian integral defined by means of the arithmetic (\ref{X1})--(\ref{X4}).

Then, what about the Bell inequality?

Of course, it is {\it not\/} violated by (\ref{46'}) despite the exact quantum mechanical form of the probability, and there is nothing paradoxical about this statement. Just try to derive any form of a Bell-type inequality for such a non-Newtonian local hidden-variable model. For example, following the steps of the Clauser-Horne reasoning one arrives at the projective-arithmetic generalization of  the Clauser-Horne inequality,
\be
0\leq 3\odot_\mathbb{X} P_{1_10_2}(\theta)\ominus_\mathbb{X} P_{1_10_2}(3\theta)\leq 1.\label{CHi'}
\ee
Inserting singlet-state probabilities into (\ref{CHi'}) one finds
\be
3\odot_\mathbb{X} P_{1_10_2}(\theta)\ominus_\mathbb{X} P_{1_10_2}(3\theta)=1
\ee
for any $\theta$, so there is no contradiction. 

The inequality that will be indeed violated is
\be
0\leq 3 P_{1_10_2}(\theta)- P_{1_10_2}(3\theta)\leq 1,\label{CHi}
\ee
but it is derived under the {\it wrong\/} assumption of additivity (\ref{1}), that does not hold for this  concrete model of non-Newtonian integration. Standard Clauser-Horne inequality (\ref{CHi}) cannot be proved for non-Newtonian hidden variables in question, so it is no surprise it is not satisfied in our model.

The readers should keep in mind that although (\ref{18}) and (\ref{19}) are simultaneously valid, this is no longer true for arbitrary linear combinations of probabilities, in particular those occurring in (\ref{CHi'}) and (\ref{CHi}).

We will now show that the relation between $p$ and $\tilde p= g_\mathbb{X}(p)$, which is at the core of the Bell inequality violation, is in fact a very special case of an infinite hierarchy of relations, based on an infinite hierarchy of arithmetics and calculi. What we intuitively regard as the `normal' or `our' arithmetic and calculus can correspond to  {\it any\/} level of the hierarchy. 

This will lead us to the notion of a Copernican hierarchy of models. 
We call them Copernican because they deprive our human point of view of the aura of uniqueness. Each level of such a hierarchy can be our level.

The standard Bell theorem describes a relation between any two neighboring levels of the hierarchy. A surprising consequence of this relation is that in the same way that Bell proved the non-existence of EPR elements of reality, it is possible to prove the non-existence of ourselves.

Well, at least the author of this paper exists as an element of reality.

\section{Copernican hierarchies}

Functions $g$ that satisfy Lemma~1 form an interesting  structure, closed under composition of maps \cite{MCKN}.

{\it Lemma 3\/}: Consider two functions $g_j:[0,1]\to [0,1]$, $j=1,2$, that satisfy assumptions of Lemma~1,
\be
g_j(p)=\frac{1}{2} + h_j\left(p-\frac{1}{2}\right),
\ee
where $h_j(-x)=-h_j(x)$. 
Then $g_{12}=g_1\circ g_2$ also satisfies Lemma~1 with $h_{12}=h_1\circ h_2$,
\be
g_{12}(p)=\frac{1}{2} + h_{12}\left(p-\frac{1}{2}\right).
\ee
Accordingly,
\be 
g_{12}(p)+g_{12}(1-p)=1
\ee
for any $p\in[0,1]$.

{\it Lemma 4\/}: Let $g^k=g\circ \dots \circ g$, $g^{-k}=g^{-1}\circ \dots \circ g^{-1}$ ($k$ times), $g^0(x)=x$. If $g$ satisfies Lemma~1, 
\be
g(p)=\frac{1}{2} + h\left(p-\frac{1}{2}\right),
\ee
then $g^k$ also satisfies Lemma~1 for any $k\in\mathbb{Z}$,
\be
g^k(p)=\frac{1}{2} + h^k\left(p-\frac{1}{2}\right).
\ee
Accordingly,
\be 
g^k(p)+g^k(1-p)=1\label{57a}
\ee
for any $p\in[0,1]$, and any integer $k$. In particular
\be 
g^{-1}(p)+g^{-1}(1-p)=1.\label{58}
\ee
The proofs are straightforward \cite{MCKN}. 

\begin{figure}
\includegraphics[width=8 cm]{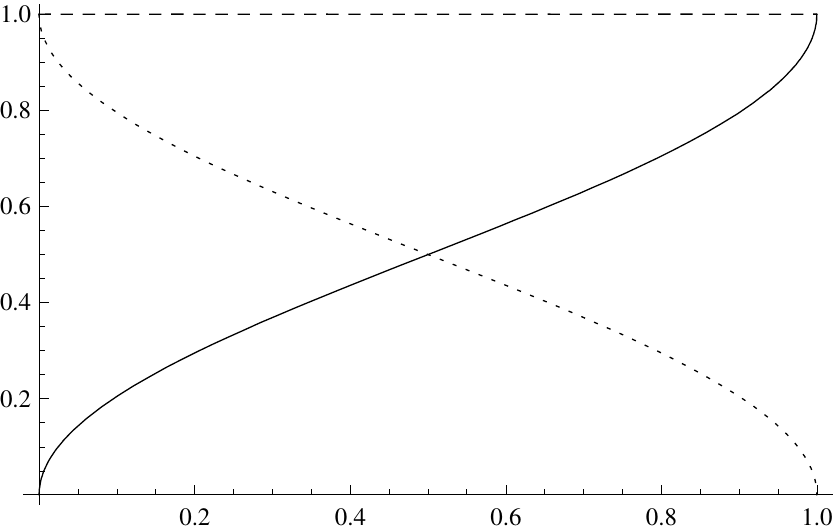}
\caption{$g^{-1}(p)=\frac{2}{\pi}\arcsin \sqrt{p}$  (full), $g^{-1}(1-p)$ (dotted), and their sum (dashed).}
\label{Fig arc}
\end{figure}

As  an illustration consider again $g(p)=\sin^2\frac{\pi}{2}p$ and
\be
g^2(p) &=&\sin^2\frac{\pi}{2}\left(\sin^2\frac{\pi}{2}p\right),\label{59}\\
g^{-1}(p) &=& \frac{2}{\pi}\arcsin \sqrt{p}.\label{60}
\ee
The cross-check of (\ref{57a}) for (\ref{59}) is simple but instructive:
\be
&{}&
g^2(p)+g^2(1-p)
\nonumber\\
&{}&\pp=
=
\sin^2\frac{\pi}{2}\left(\sin^2\frac{\pi}{2}p\right)
+
\sin^2\frac{\pi}{2}\left(\sin^2\frac{\pi}{2}(1-p)\right)
\nonumber\\
&{}&\pp=
=
\sin^2\frac{\pi}{2}\left(\sin^2\frac{\pi}{2}p\right)
+
\sin^2\frac{\pi}{2}\left(\cos^2\frac{\pi}{2}p\right)
\nonumber\\
&{}&\pp=
=
\sin^2\frac{\pi}{2}\left(\sin^2\frac{\pi}{2}p\right)
+
\sin^2\frac{\pi}{2}\left(1-\sin^2\frac{\pi}{2}p\right)
\nonumber\\
&{}&\pp=
=
\sin^2\frac{\pi}{2}\left(\sin^2\frac{\pi}{2}p\right)
+
\cos^2\frac{\pi}{2}\left(\sin^2\frac{\pi}{2}p\right)
=1.\nonumber
\ee
An analogous proof for (\ref{60}) is left as an exercise. Fig.~\ref{Fig arc} shows the result.

We are inclined to believe that `our' arithmetic corresponds to $k=0$.
So, consider any binary probabilities from level 0,
\be
p_0+p_1 &=& 1,\label{61}
\ee
and those from level $k$,
\be
g^k(p_0)+g^k(p_1)&=&1.\label{62}
\ee
Denoting $P_0=g^k(p_0)$, $P_1=g^k(p_1)$, we obtain a symmetric rule,
\be
P_0+P_1 &=& 1,\label{63}\\
g^{-k}(P_0)+g^{-k}(P_1)&=&1.\label{64}
\ee
In both cases $k$ is an arbitrary integer: positive, negative, or zero. 

The question is: How do we know that it is  $p_A$ and not $P_A$ that defines level-zero probabilities?

We can phrase the same question in arithmetic terms.
To this end, assume $g(p)$ is a restriction to $[0,1]$ of some bijection $g_\mathbb{R}:\mathbb{R}\to \mathbb{R}$, that satisfies 
$g_\mathbb{R}(1)=g(1)=1$. We can act on both sides of (\ref{62}) with $g^{-k}$, while  on both sides of (\ref{64}) with $g^k$, obtaining
\be
g^{-k}\left(g^k(p_0)+g^k(p_1)\right)=p_0\oplus_k p_1 =1,\label{62'}
\ee
and
\be
g^k\left(g^{-k}(P_0)+g^{-k}(P_1)\right)=P_0\oplus_{-k}P_1=1.\label{64'}
\ee
We have no criterion that could tell us which of the four additions --- (\ref{61}), (\ref{63}), (\ref{62'}), or (\ref{64'}) --- defines the level of description  we employ in everyday life. Which of these two probabilities, and which of the several ways of adding them,  is our usual way of processing experimental data?

Which of the three additions, $+$, $\oplus_{k}$, or $\oplus_{-k}$, is the one we have learned as kids? 

Which of the three derivatives: (\ref{28}), or
\be
\frac{{\rm D} F(x)}{{\rm D}x}
=
\lim_{\delta\to 0}
\big(F(x\oplus_{k}\delta_{k})\ominus_{k}F(x)\big)\oslash_{k}\delta_{k},
\label{DFk}
\ee
or
\be
\frac{{\rm D} F(x)}{{\rm D}x}
=
\lim_{\delta\to 0}
\big(F(x\oplus_{-k}\delta_{-k})\ominus_{-k}F(x)\big)\oslash_{-k}\delta_{-k},
\label{DF-k}
\ee
is the one we have mastered during our undergraduate education?

Last but not least, which of the three integrals, $\int F(x){\rm D}x$, is the one that should define a hidden-variable theory?

Numerous fundamental answers are possible.

One possibility is that Nature prefers only one $k\in\mathbb{Z}$ as the true physical arithmetic with some fixed form of $g$, determined by some unknown physical law. This is the situation we encounter in special relativity when we add velocities by means of $g=\tanh$. In principle, we can detect such a physical $g$ in experiment. In \cite{MCdark2} it is shown that problems with dark energy may indicate that time at cosmological scales involves a nontrivial $g\sim \sinh$. If this conclusion were true, dark energy would be as unreal  as the luminiferous aether.  

The second fundamental possibility is that all these possibilities are simultaneously true. Perhaps there is no preferred $k$, like there is no preferred rest frame, or preferred point of observation of the Universe. Only relative $k$ might be observable. Such an option is intriguing and tempting from a theoretical perspective. It could mean, for example, that the same physical law might have its mathematical representations at any level of the hierarchy, and each of these representations might be meaningful. Violation of Bell-type inequalities would then be a conflict of predictions derived at level $k$, but tested at level $k+1$.

A similar conflict occurs if Bob concludes that Alice will never reach the Schwarzschild radius, and yet she crosses it in a finite time.

It remains to say something about the conflicts that occur between non-neighboring levels of the hierarchy. 
We will see that other well known bounds, such as the Tsirelson inequality characterizing Hilbert-space models of probability, can be easily circumvented as well. 

\section{Beyond Tsirelson bounds}

The standard Clauser-Horne inequality (\ref{CHi})
\be
0\leq 3 P_{1_10_2}(\theta)- P_{1_10_2}(3\theta)\leq 1,
\ee
is derived for joint probabilities limited by 
\be
0\le P_{1_10_2}(\alpha)\le 0.5.
\ee
The  absolute bounds for such a linear combination of probabilities are therefore 
\be
-0.5\leq 3 P_{1_10_2}(\theta)- P_{1_10_2}(3\theta)\leq 1.5.
\ee
Tsirelson bounds are narrower,
\be
-\frac{\sqrt{2}-1}{2}
\leq 3 P_{1_10_2}(\theta)- P_{1_10_2}(3\theta)\leq 
\frac{\sqrt{2}+1}{2}.
\ee
In order to understand the influence of $k$ on violation of  Clauser-Horne $k=0$ inequalities we have to estimate the expression \cite{MCKN}
\be
X(g^k,\theta)
=
3g^k\left(\frac{\theta}{2\pi}\right)
-
g^k\left(\frac{3\theta}{2\pi}\right),
\ee
where $0\le \theta <\pi/3$. The singlet-state example corresponds in this range of parameters to $g^1(p)=\frac{1}{2}\sin^2 (\pi p)$. For  $\theta=\pi/4$ we find
\be
X(g^1,\pi/4)
&=&
\frac{3}{2}\sin^2 \left(\pi \frac{\pi/4}{2\pi}\right)
-
\frac{1}{2}\sin^2 \left(\pi \frac{3\pi/4}{2\pi}\right)
\nonumber\\
&=&
-\frac{\sqrt{2}-1}{2}=-0.20711,
\ee
that is, the maximal left Tsirelson bound. This is what is usually called the maximal (left) violation of the Clauser-Horne inequality by singlet-state quantum probabilities. For other values of $k$ we find:
\be
X(g^0,\pi/4) &=& 0,\label{Xmin0}\\
X(g^1,\pi/4) &=& -0.20711,\\
X(g^2,\pi/4) &=& -0.39602,\\
X(g^3,\pi/4) &=& -0.48669,\\
X(g^4,\pi/4) &=& -0.49978,\\
&\vdots& \nonumber\\
X(g^\infty,\pi/4) &=& -0.5.\label{Xmin5}
\ee

Of course, as stressed before, the choice of $k=0$ as the reference level is arbitrary. Level $k=2022$ probabilities violate level $k=2021$  inequalities in exactly the same way as quantum mechanics violates the standard Clauser-Horne inequality.

More importantly, $k=2$ probabilities violate $k=1$ inequalities in the same way as $k=1$ probabilities violate $k=0$ inequalities. If $k=0$ elements of reality do not exist, then $k=1$ elements of reality do not exist either. Accepting the logic of Bell's theorem, can we prove by induction that nothing exists?

Slightly modifying the experimental configuration one obtains the maximal right violations. In our formalism, the function to estimate is
\be
Y(g^k,\theta)
=
3g^k\left(\frac{3\theta}{2\pi}\right)
-
g^k\left(\frac{\theta}{2\pi}\right).
\ee
We find:
\be
Y(g^0,\pi/4) &=& 1,\label{Xmax0}\\
Y(g^1,\pi/4) &=& (\sqrt{2}+1)/2=1.20711,\\
Y(g^2,\pi/4) &=& 1.39602,\\
Y(g^3,\pi/4) &=& 1.48669,\\
Y(g^4,\pi/4) &=& 1.49978,\\
&\vdots& \nonumber\\
Y(g^\infty,\pi/4) &=& 1.5.\label{Xmax5}
\ee
All these models are local-realistic, observers have free will, detectors are ideal... The only modification is in the presence of the bijection $g$ that links arithmetics, calculi, and probabilities at various levels of the hierarchy. 

Our $g^k$ plays a role analogous to $g_r$ that linked experimental data collected at different neighborhoods of  a collapsing star. 

Both examples are based on principles of relativity. We have learned to live with special relativity, general relativity, and the Copernican principle. 

It is time to learn to live with  the arithmetic principle of relativity.

\begin{figure}
\includegraphics[width=7.3 cm]{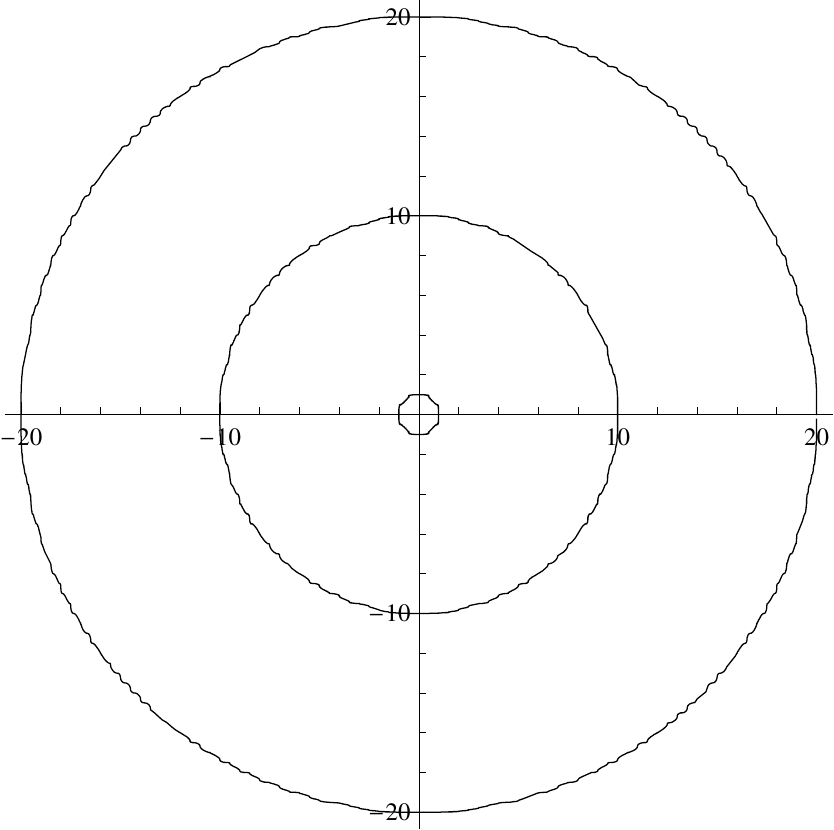}
\includegraphics[width=7.3 cm]{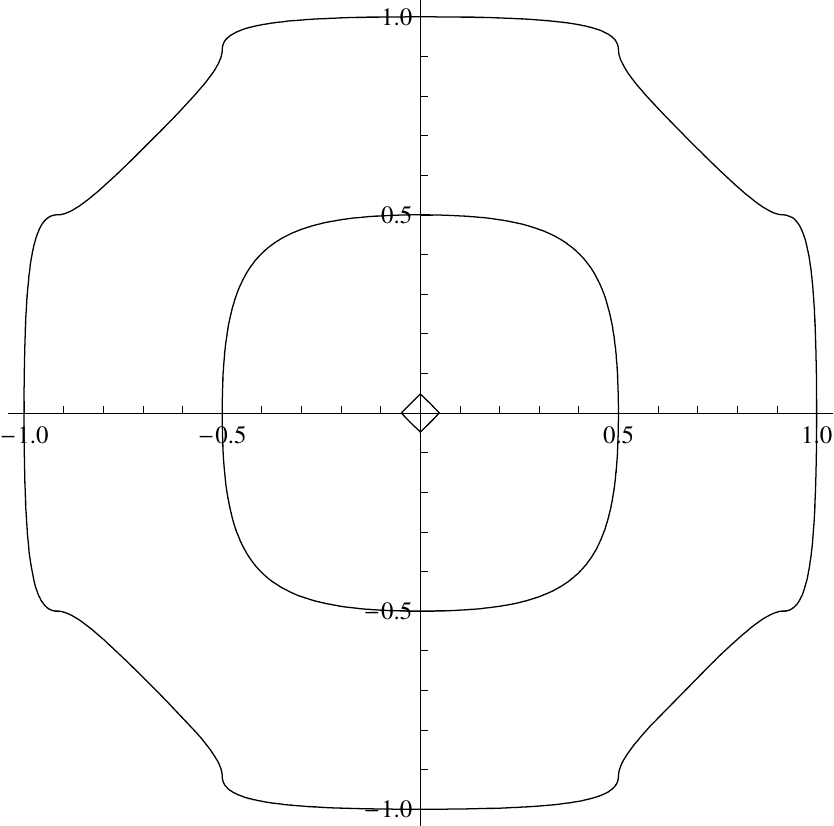}
\includegraphics[width=7.3 cm]{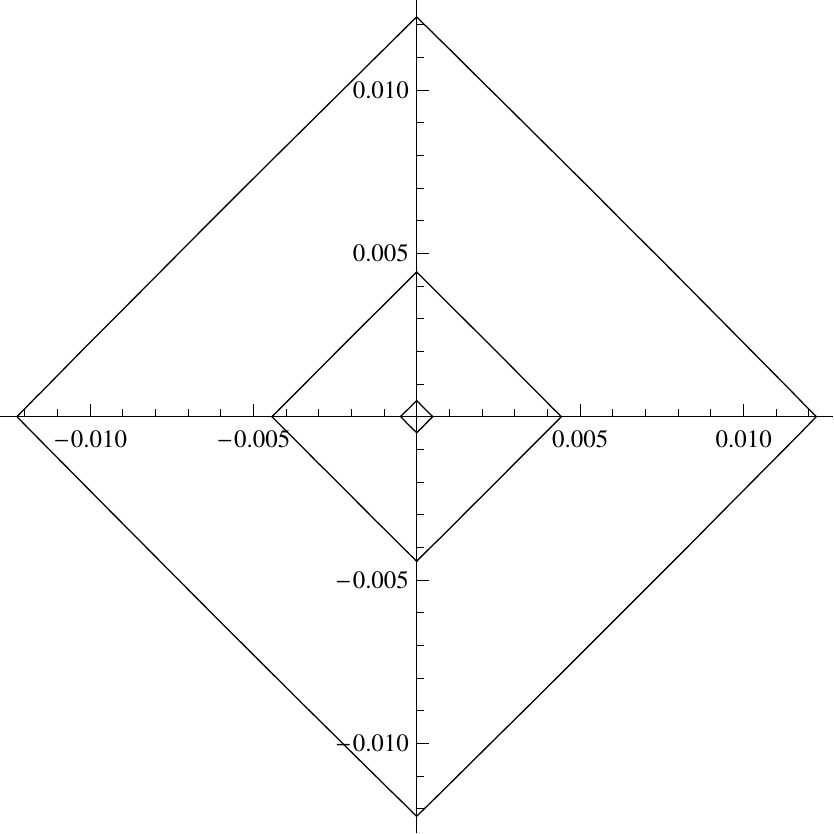}
\caption{Seven circles of different radii described by (\ref{circle}) with the arithmetic $\{\mathbb{R},\oplus_1,\ominus_1,\odot_1,\oslash_1\}$ defined by (\ref{f^-1}) and (\ref{f}). Despite appearances, all these circles are rotationally invariant. Lower illustrations are the close-ups of the upper ones.}
\label{Fig3}
\end{figure}

\section{Correspondence principles}
\label{Correspondence principles}
Trigonometric functions $\cos_\mathbb{X}: \mathbb{X}\to \mathbb{X}$, $\sin_\mathbb{X} :\mathbb{X}\to \mathbb{X}$, are defined by
\be
\cos_\mathbb{X} x &=& g_\mathbb{X}\big(\cos f_\mathbb{X}(x)\big),\\
\sin_\mathbb{X} x &=& g_\mathbb{X}\big(\sin f_\mathbb{X}(x)\big).
\ee
They satisfy all the usual trigonometric relations, of course with respect to appropriate arithmetic operations. They also satisfy all the usual differential relations, of course with respect to appropriate non-Newtonian derivatives. In particular, they define circles by 
\be
\theta\mapsto (r\odot_\mathbb{X} \cos_\mathbb{X} \theta,r\odot_\mathbb{X}\sin_\mathbb{X}  \theta).\label{circle}
\ee
Let us now take the bijections (\ref{f^-1}), (\ref{f}), we have used to reconstruct singlet state probabilities. Fig.~\ref{Fig3} shows seven circles defined by (\ref{circle}) for decreasing radii. A lower picture is a close-up of its upper neighbor.  All the circles are given by the same formula (\ref{circle}), with the same bijection $g_\mathbb{X}$. The greater the radius, the more circular the shape. Simply put, the larger the $x$ argument, the more difficult it is to tell $g_\mathbb {X}(x)$ from $x$. However, the readers must bear in mind that all these circles are {\it truly\/} rotationally invariant! They have been generated as homogeneous spaces of the rotation group in 2D --- the only nonstandard element being the choice of arithmetic. 

The notion of a {\it hidden\/} or {\it internal\/} symmetry, often used in particle physics, seems especially adequate here. Each of these circles would have looked `normal' if we had reprogrammed Wolfram Mathematica to make the plots in the arithmetic $\{\mathbb{R},\oplus_1,\ominus_1,\odot_1,\oslash_1\}$.

The limit $r\to \infty$ plays a role of a correspondence principle with the ordinary, rotational external symmetry. 
The obvious similarity to the classical limit of quantum mechanics is striking. Other examples of arithmetic correspondence principles can be found in \cite{MC2015} and \cite{Pilat2022}. An analogous correspondence principle occurs in idempotent analysis \cite{LMS}.

\section{Implications for cryptography}

In 1862, more than a century before Bell's paper, George Boole submitted to Philosophical Transactions of the Royal Society the article {\it On the theory of probabilities\/}, where he introduced inequalities imposing constraints on our `possible experience' \cite{Boole}. Three decades after Bell's theorem, in 1994, Itamar Pitowsky noticed that Boole's inequalities are inequalities of a Bell type \cite{Pitowsky1994}. 

Boole inequalities  defined `possible experience' in common-sense categories appropriate for 1862. 
Boole's scientific paradigm has been falsified by quantum mechanics.

If someone had asked Boole if he could give an example of a system that violates his inequalities, he probably would have answered in the negative. Treating his negative answer as the ultimate proof that Nature has to comply with Boole inequalities, we would prove that  quantum mechanics is logically  impossible.

Bell inequalities  defined `possible experience' in common-sense categories appropriate for 1964. 
Grossman and Katz book appeared in 1972, but its implications for Bell's theorem went unnoticed until very recently.

In light of these results, what is the actual status of all the claims about fundamental security of quantum cryptography \cite{BB,Ekert,BBM92,Gisin,ACP}? We typically base them on the belief that EPR elements of reality cannot exist. Protocols that are not based on a hidden-variable argumentation (such as the Bennett-Brassard-Mermin one \cite{BBM92}, essentially based on rotational invariance of singlet-state probabilities)  {\it can\/} be  successfully attacked in non-Newtonian local hidden variable theories: Non-Newtonian hidden variables are rotationally invariant, because the rotation group works there by means of the {\it hidden\/} representation depicted in Fig.~3.

Furthermore, what if the Newtonian paradigm of contemporary quantum mechanics will be falsified one day by some new theory? 

What if it {\it is\/} already falsified?
What if our enemies, whoever they may be, are well ahead of us and know systems that can mimic quantum probabilities by means of non-Newtonian hidden variables?
Can they hack entangled-state  quantum communication channels? 

Can a no-go theorem, based on algebraic rather than probabilistic properties of quantum mechanics, cure the arithmetic loophole in quantum proofs of security?

How to guarantee that we are not in the position of German cryptographers in the 1930s, so happy with their Enigma and its  security certified  by appropriate theorems, while at the same time it was systematically hacked by the Polish Cypher Bureau?

The list of open questions is longer. 

\section{Non-Newtonian Quantum Mechanics}

Non-Newtonian hidden variables are not meant as an alternative to quantum mechanics. 

However, non-Newtonian calculus paves a way to natural generalizations of quantum mechanics (quantum mechanics on a Cantor set is an example \cite{MCdark2}). 
The resulting theory is non-Newtonian linear but Newtonian non-linear. Such a form of non-linear quantum mechanics \cite{IBB,Zeilinger,Weinberg,DG} is isomorphic to the standard textbook theory, so is free of all the difficulties that have plagued the formalisms based on non-linear Schr\"odinger equations
\cite{Gisin1989,Polchinski,MC1991,MCHDD}.

Yet, `mathematically isomorphic' is not synonymous to `physically equivalent'. 

The following three examples illustrate the idea.

\begin{figure}
\includegraphics[width=7 cm]{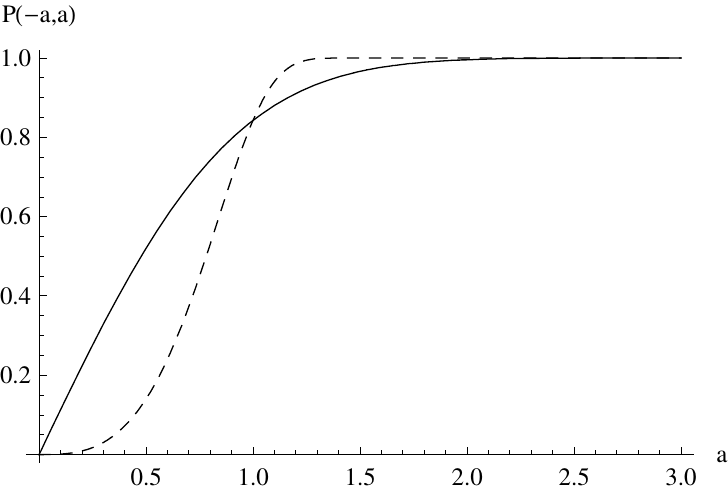}
\caption{Probability (\ref{97}) of finding a particle in $[-a,a]$ for $0\leq a\leq 3$, with $\tilde\psi(r)\sim e^{-r^2/2}$ representing the ground state of a quantum harmonic oscillator, in dimensionless units, $\tilde U(r)=r^2$, for (a) the ordinary arithmetic (full), and (b) projective arithmetics in ${\mathbb X}={\mathbb R}={\mathbb Y}$  defined by $f_\mathbb{X}(x)=x^3$, $f_\mathbb{Y}(x)=x$ (dashed).  The ordinary arithmetic is experimentally distinguishable from the projective one because limits of integration are $[-a^3,a^3]$ instead of the usual $[-a,a]$.}
\label{Fig erf}
\end{figure}
\begin{figure}\center
\includegraphics[width=7 cm]{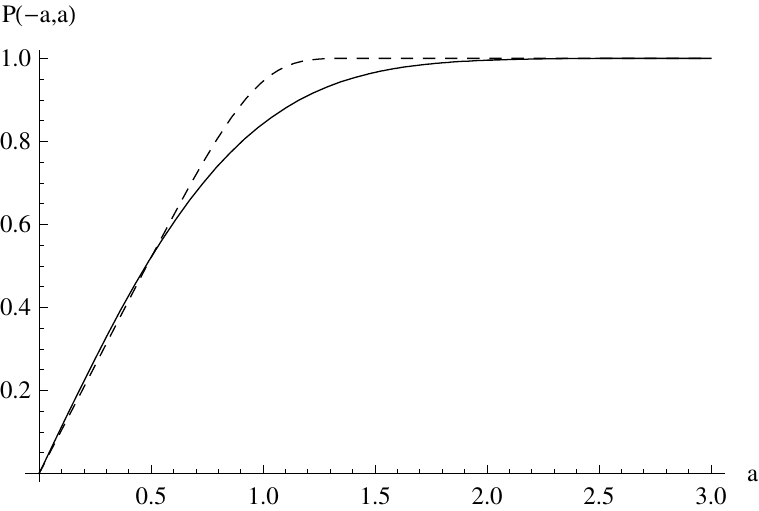}
\caption{The same situation as in the previous figure, but now with $f_\mathbb{X}(x)=x^3=f_\mathbb{Y}(x)$.}
\label{Fig erf1}
\end{figure}

Assume $\mathbb{X}=\mathbb{Y}=\mathbb{R}$ with projective arithmetics defined by some $f_\mathbb{X}:\mathbb{R}\to \mathbb{R}$ and $f_\mathbb{Y}:\mathbb{R}\to \mathbb{R}$.
Let $\psi:\mathbb{X}\to\mathbb{Y}$ be a solution of
\be
H\psi(x)
=
\ominus_\mathbb{Y}\psi''(x)\oplus_\mathbb{Y} U(x)\odot_\mathbb{Y}\psi(x)
=
E\odot_\mathbb{Y}\psi(x)
,
\ee
where $\psi''(x)$ is the non-Newtonian second derivative.
Normalization of states is assumed in the form 
\be
\langle\psi|\psi\rangle
=
\int_{(-\infty)_\mathbb{X}}^{\infty_\mathbb{X}}|\psi(x)|^{2_\mathbb{Y}}{\textrm D}x
=
1_\mathbb{Y}=g_\mathbb{Y}(1).
\ee
(For real-valued $\psi(x)$ the modulus in 
\be
|\psi(x)|^{2_\mathbb{Y}}=\psi(x)\odot_\mathbb{Y}\psi(x)
\ee
is redundant, but we keep it to make the notation less awkward.) 
Probability of finding a particle in $[a,b]\subset \mathbb{X}$ equals
\be
P(a,b)
=
\int_{a}^{b}|\psi(x)|^{2_\mathbb{Y}}{\textrm D}x.\label{94}
\ee
As usual, $\psi=g_\mathbb{Y}\circ \tilde\psi\circ f_\mathbb{X}$, $U=g_\mathbb{Y}\circ \tilde U\circ f_\mathbb{X}$ (compare (\ref{33})). Let  
$\tilde \psi''\big(f_\mathbb{X}(x)\big)$ be the Newtonian second derivative of $\tilde\psi$ with respect to $f_\mathbb{X}(x)$, so that the non-Newtonian Schr\"odinger equation is equivalent to the usual Newtonian equation,
but with redefined parameters:
\be
f_\mathbb{Y}(E)\tilde \psi(r)
&=&
-\tilde \psi''(r)+ \tilde U(r)\tilde \psi(r),\\
1 &=&
\langle\tilde\psi|\tilde\psi\rangle
=
\int_{-\infty}^{\infty}|\tilde\psi(r)|^{2}{\textrm d}r.
\ee
Now let us consider   $f_\mathbb{X}(x)=x^3$, $f_\mathbb{Y}(x)=x$. Then 
$\psi=\tilde \psi\circ f_\mathbb{X}$, 
$U=\tilde U\circ f_\mathbb{X}$, and the Schr\"odinger equation is just
\be
E \tilde\psi(r)
&=&
-\tilde\psi''(r)+ \tilde U(r)\tilde\psi(r),
\ee
so apparently the problem is completely equivalent to the standard one. 
However, due to triviality of $f_\mathbb{Y}$ and non-triviality of $f_\mathbb{X}$, probability (\ref{94})  is now explicitly given by
\be
P(a,b)
&=&
\int_{f_{X}(a)}^{f_{X}(b)}|\tilde\psi(r)|^{2}{\textrm d}r
=
\int_{a^3}^{b^3}|\tilde\psi(r)|^{2}{\textrm d}r.\label{97}
\ee
As we can see, despite mathematical banality of the problem, the non-Diophantine arithmetic of $\mathbb{X}$ does influence the probability of finding the particle in the interval $[a,b]$ because the integral is over $[a^3,b^3]$. Fig.~\ref{Fig erf} shows the probability of finding the particle in $[-a,a]$ as a function of $a$.

Taking $f_\mathbb{X}(x)=x^3=f_\mathbb{Y}(x)$ we obtain probability depicted in Fig.~\ref{Fig erf1}.

As the third example consider $f_\mathbb{X}(x)=x$, $f_\mathbb{Y}(x)=x/\sqrt{|x|}$, and $g_\mathbb{Y}(x)=f^{-1}_\mathbb{Y}(x)=x^3/|x|$. Now,
\be
P(a,b)
&=&
\left(\int_{a}^{b}|\tilde \psi(r)|^{2}{\textrm d}r\right)^2.
\ee
Projective  addition of probabilities looks here like a superposition principle from quantum mechanics,
\be
P(a,c) &=&P(a,b)\oplus_\mathbb{Y}P(b,c)\nonumber\\
&=&
\left(\sqrt{P(a,b)}+\sqrt{P(b,c)}\right)^2.
\ee

Theories based on non-Newtonian calculi involve the same physical principles, but their mathematical forms may differ between one another.

Is there any natural law that determines the form of arithmetic and calculus? 

\section{Towards a new paradigm}

Paul Benioff, a pioneer of quantum computation, was among those physicists who believed that physics and mathematics should be logically formulated at a unified level \cite{B2002,B2005}. According to Benioff, physical or geometric quantities do not possess numerical values per se, but these values are introduced through `value maps'. Natural numbers are elements of any well-ordered set, and in themselves do not possess numerical values. A value map takes a number and turns it into an object with concrete numerical properties. This is somewhat similar to the idea that `zero', the neutral element of addition, can be in fact an arbitrary point $0_\mathbb{X}\in \mathbb{X}$ provided $\mathbb{X}$ can be bijectively mapped onto $\mathbb{R}$ by means of some $f_\mathbb{X}$ that fulfills $f_\mathbb{X}(0_\mathbb{X})=0\in \mathbb{R}$. Benioff considered only linear value maps, but allowed for the possibility of value-map fields. One of his conclusions was that scalar value maps are in many respects analogous to the Higgs field \cite{B2015,B2016a,B2016b}.

A fundamental role of the set of bijections occurs also in Etesi's recent reformulation of black-hole entropy \cite{Etesi}. The `arithmetic continuum' $\mathbb{R}$ plays there a role of a gas subject to thermodynamic laws while the black hole entropy is a purely set-theoretic notion, related to G\"odel's first incompleteness theorem.

The approach advocated in our work involves arithmetics and calculi constructed by means of bijections that can be regarded either as (global) non-linear Benioff's maps, or compositions of value maps with some bijections. More precisely, the `nonlinear' bijection $f_\mathbb{X}: \mathbb{X}\to \mathbb{R}$ is here always linear but with respect to $\oplus_\mathbb{X}$, $\ominus_\mathbb{X}$, $\odot_\mathbb{X}$ and $\oslash_\mathbb{X}$. The non-Newtonian formalism in its most general form demands only bijectivity of $f_\mathbb{X}$. One does not impose continuity or topological conditions on either $f_\mathbb{X}$ or $\mathbb{X}$. $f_\mathbb{X}$ is always smooth in the topology and calculus it induces from $\mathbb{R}$, even if $\mathbb{X}$ is as weird as Cantor or Sierpi\'nski fractals. Non-Newtonian derivatives of $f_\mathbb{X}$ and $g_\mathbb{X}$ are `trivial' (equal to 1 and $1_\mathbb{X}$, respectively \cite{BC}), because from the point of view of the projective arithmetic in $\mathbb{X}$ the map $f_\mathbb{X}$ behaves as the identity map. 

The duality between non-Newtonian linearity and Newtonian non-linearity is one of the trademarks of the new paradigm. This is not the usual  linearization of a nonlinear problem by a nonlinear change of variables. The  idea  can be traced back to  Maslov's superposition principle and its application to nonlinear optimization problems \cite{Maslov1}.

The resulting structure is incredibly flexible. It automatically leads to well-behaved calculi on all sets whose cardinalities are the same as the cardinality of the continuum. The resulting relativity principle (relativity of arithmetic and calculus) is much more general than principle of general covariance. 

Non-Newtonian calculus has huge potential of unification and systematization of various ideas scattered over mathematical and physical literature \cite{Entropy}. It is quite typical, however, that even if some elements of non-Newtonian thinking can be identified in those works, their arithmetic aspects are not exploited in their full generality. 

For example, velocities in special relativity are added and subtracted in the projective way, but it is difficult to find a paper where repeated addition would be replaced by multiplication and its inverse, division. I have found only one place in relativistic physics where velocity $v=c\tanh 1$, the `one' in special-relativistic projective multiplication,  plays a distinguished role \cite{MC2022}.  Kolmogorov-Nagumo averages \cite{KN1,KN2}, the departure point of R\'enyi's studies on generalized entropies \cite{Renyi}, are exactly the averages in the sense of projective arithmetic. However, when R\'enyi discussed additivity of his $\alpha$-entropies, he did not think of additivity in the same sense as the one he implicitly used in Kolmogorov-Nagumo averaging. Various forms of projective arithmetic operations and derivatives have been studied in the context of generalized statistical physics and thermodynamics by Kaniadakis \cite{K1a,K1b,K2,K3}, but only some of the derivatives he invented were non-Newtonian, whereas the others were neither Newtonian nor non-Newtonian, a fact explaining why only the non-Newtonian ones have found applications \cite{Entropy}. The whole field of psychophysics is implicitly based on projective addition (cf. Chapter 7 in \cite{BC}). Typically, we are unaware that decibels and star magnitudes  correspond to logarithmic scales because our sensory systems induce a projective arithmetic in our brains, based on approximately logarithmic bijections (the Weber-Fechner law). However, although projective addition is here essential, the remaining three arithmetic operations are not employed. Certain elements of non-Newtonian integral calculus are present in cepstral analysis and homomorphic filtering of images \cite{cep}. Fractional derivatives can be regarded as non-Newtonian first derivatives, but only when formulated in the so called $F^\alpha$ formalism \cite{Alireza}. Fuzzy calculus is non-additive but not-necessarily fully non-Newtonian, and this is why the fundamental theorem of calculus does not necessarily work. Non-additive probability, somewhat similar to our non-Newtonian hidden variables (as based on non-additive Vitali and Choquet  integrals) is a standard element of modern decision theory \cite{Schmeidler,Marinacci}.

Perhaps the most radical view on generalized arithmetics is due to Mark Burgin who studied arithmetics that are {\it not\/} isomorphic to the aritmetic of natural numbers \cite{Burgin77}. One of his goals was to replace inconsistent arithmetics (eg. the computer aritmetic based on the  notion of `machine infinity': $\infty_M<\infty$, $\infty_M+1=\infty_M$)
by arithmetics that are consistent but non-Diophantine. 

Similarly radical is the approach of Sergeyev \cite{Sergeyev}, where infinities and infinitesimals are reformulated in a more intuitive and, essentially,   non-Diophantine way. Here infinity of integers is twice bigger than infinity of  natural numbers, while events of zero probability cannot happen (as opposed to the Kolmogorovian formalism based on measures) \cite{Calaude}. Such a new arithmetic and probabilistic paradigm often  turns out to be more practical than the usual Kolmogorovian framework, just to mention Sergeyev's Infinity Calculator software. 

Interference of probabilities is one of the greatest puzzles of quantum mechanics. Quantum probabilities sometimes behave as if they were negative, a situation known from projective arithmetics based on, say, $g_\mathbb{X}(p)=\ln p$. Risk aversion paradoxes in economics can be modeled by non-additive integrals \cite{Schmeidler}, but a new tendency appears where the same effects are modeled by quantum probabilities \cite{AST}. Ironically, as here we have shown that quantum probabilities can be classical but non-Newtonian, some authors start to replace non-additive integrals from classical economics by quantum probabilities \cite{Bruza}. 

Another aspect of interference is the superposition principle and the problem of linearity of quantum mechanics. Non-Newtonian linear Schr\"odinger equation can be Newtonian non-linear. In such nonlinear quantum mechanics the superposition principle remains the same as in the linear theory, only the meaning of `plus' and `times' is different. The same type of duality was introduced by Maslov to optimization theory \cite{Maslov1}, with the key idea that something very difficult in a non-linear framework can become easy if we rewrite the problem in a new arithmetic. 

Speaking of a non-Newtonian paradigm one typically has in mind a non-Newtonian theory of gravity (hence general relativity), or non-Newtonian mechanics (hence quantum mechanics). In the new paradigm that looms on the horizon, the term non-Newtonian may be understood in a much broader sense.

\acknowledgments

I'm indebted to Kamil Nalikowski for discussions. Calculations were carried out at the Academic Computer Center in Gda{\'n}sk. The work was supported by the CI TASK grant `Non-Newtonian calculus with interdisciplinary applications'.

\end{document}